\documentclass[
 aps,%
 prl,%
 reprint, twocolumn,%
 groupedaddress,%
 superscriptaddress,%
  showpacs,%
 letterpaper,%
 amsfonts,%
 footinbib,%
 10pt,
 floatfix,%
 noeprint%
]{revtex4-1}
\RequirePackage{amsmath,amssymb,bm}
\RequirePackage{graphicx}

\RequirePackage{hyperref}
\hypersetup{%
  breaklinks = {true},
  citecolor = {blue},
  colorlinks = {true},
  linkcolor = {blue},
  urlcolor = {Magenta},
  pdfauthor = {Vitor M. Pereira},
  pdfcreator = {\LaTeX\ and \flqq hyperref\frqq},
}

\graphicspath{ {./Figs/} } 
\newcommand{\figmaxwidth}{\columnwidth}

\RequirePackage[usenames,dvipsnames]{xcolor}

\newcommand{\Fref}[1]{Fig.~\ref{#1}}
\newcommand{\Frefs}[1]{Figs.~\ref{#1}}

\newcommand{\teq}{{\,=\,}}
\newcommand{\tequiv}{{\,\equiv\,}}
\newcommand{\ttimes}{{\,\times\,}}

\newcommand{\tsim}{{\,\sim\,}}
\newcommand{\tplus}{{\,+\,}}
\newcommand{\tminus}{{\,-\,}}
\newcommand{\tapprox}{{\,\approx\,}}
\newcommand{\tne}{{\,\ne\,}}
\newcommand{\etal}{\emph{et al.}}

\newcommand{\Tc}{T_{\text{c}}}
\newcommand{\Tsc}{T_\text{sc}}
\newcommand{\Ef}{E_{\text{F}}}
\newcommand{\Qcdw}{\mathbf{Q}_\text{cdw}}
\newcommand{\ba}{\mathbf{a}}
\newcommand{\bb}{\mathbf{b}}
\newcommand{\bc}{\mathbf{c}}
\newcommand{\bk}{\bm{k}}
\newcommand{\bM}{\bm{M}}
\newcommand{\bq}{\bm{q}}

\newcommand{\ve}{\varepsilon}

\newcommand{\TiSe}{TiSe$_2$}
\newcommand{\CuTiSe}{Cu$_x$TiSe$_2$}
\newcommand{\SI}{supplementary information}
\newcommand{\SFref}[1]{Fig.~S{#1}}

\usepackage{pdfpages}
\makeatletter\AtBeginDocument{\let\LS@rot\@undefined}\makeatother

\begin{document}

\title{Reproduction of the charge density wave phase diagram in 
\texorpdfstring{1T-TiSe$_2$}{1T-TiSe2} exposes its excitonic character}

\author{Chuan Chen}
\affiliation{Centre for Advanced 2D Materials and Graphene Research Centre, 
National University of Singapore, Singapore 117546}  
\affiliation{Department of Physics, National University of Singapore, Singapore 
117542} 

\author{Bahadur Singh}
\affiliation{Centre for Advanced 2D Materials and Graphene Research Centre, 
National University of Singapore, Singapore 117546}
\affiliation{Department of Physics, National University of Singapore, Singapore 
117542} 

\author{Hsin Lin}
\affiliation{Institute of Physics, Academia Sinica, Taipei 11529, Taiwan}  
  
\author{Vitor M. Pereira}
\email[Corresponding author: ]{vpereira@nus.edu.sg}
\affiliation{Centre for Advanced 2D Materials and Graphene Research Centre, 
National University of Singapore, Singapore 117546}  
\affiliation{Department of Physics, National University of Singapore, Singapore 
117542} 

\date{\today}

\begin{abstract}

Recent experiments suggest that excitonic degrees of freedom play an important 
role in precipitating the charge density wave (CDW) transition in 1T-\TiSe{}.
Through systematic calculations of the electronic and phonon spectrum 
based on density functional perturbation theory, we show that the predicted 
critical doping of the CDW phase overshoots the experimental value by one order 
of magnitude. In contrast, an independent self-consistent many-body calculation 
of the excitonic order parameter and renormalized band structure is able to 
capture the experimental phase diagram in extremely good qualitative and 
quantitative agreement. 
This demonstrates that electron-electron interactions and the excitonic 
instability arising from direct electron-hole coupling are pivotal to 
accurately describe the nature of the CDW in this system. This has important 
implications to understand the emergence of superconductivity within the CDW 
phase of this and related systems.
\end{abstract}

\maketitle

The layered structure of metallic transition metal dichalcogenides (TMDs) has 
long made them archetypes to study the interplay between charge order, lattice 
instabilities and superconductivity (SC) in both quasi 
\cite{Wilson1978,Clerc2007a,Rossnagel2011,Kusmartseva2009,Morosan2006,Joe2014} 
and strictly 2D settings \cite{Chen2015,Tsen2015,Li2015,Xi2015a,Xi2016}. 
One of their common characteristics is that the SC order is stabilized 
within (sometimes deeply) a charge density wave (CDW) phase and the 
phase boundary is rather sensitive to the electronic density. 
1T-TiSe$_2$ (\TiSe, in short) is a particularly noteworthy case and will be 
our focus. It is a low-density semi-metal that undergoes a transition 
to a commensurate triple-$\bq$ CDW at a relatively high temperature that 
increases from $\Tc\tsim 200\,$K in bulk \cite{DiSalvo1976}, to about $240\,$K 
in monolayers \cite{Goli2012,Chen2015}. The ordering vectors double the unit 
cell: $\Qcdw \teq 0.5\,(\ba^*{+}\bb^*{+}\bc^*) \teq \Gamma L$ in the bulk 
\cite{DiSalvo1976}, and $\Qcdw \teq 0.5\,(\ba^*{+}\bb^*) \teq \Gamma M$ in 
the monolayer \cite{Chen2015,Fang2017}; the other two wave vectors are 
symmetric counterparts of $\Qcdw$ under $C_3$ rotations.

With no Fermi surface nesting \cite{Pillo2000} and a robust periodic lattice 
distortion (PLD) in tandem with the CDW \cite{DiSalvo1976}, it is natural to 
consider the role played by soft phonons arising from a strong and 
$k$-textured electron-phonon coupling, similarly to the cases of 
2H-NbSe$_2$ or 2H-TaSe$_2$ \cite{Moncton1975,Weber2011a,Leroux2015}. 
This would find support in density functional theory (DFT) calculations that 
reveal softening of an acoustic mode at $\Qcdw$ 
\cite{Motizuki1981,Calandra2011,Singh2017} in agreement with inelastic 
scattering experiments \cite{Holt2001,Weber2011}. 
However, despite structural similarities, \TiSe{} is a fundamentally different 
electronic system where one expects enhanced electronic interactions: The 
band structure of the normal state has small overlapping electron and hole 
pockets offset in momentum by precisely $\Qcdw$\cite{Bachrach1976,Zunger1978}, 
which strongly hints at a possible electronic instability of the excitonic 
type \cite{DiSalvo1976,Traum1978}. Despite the long-standing theoretical 
prediction for the conditions under which an excitonic insulator ground state 
should emerge \cite{Keldysh1965,Jerome1967,Kohn1967}, no representative system 
has yet been decisively found.

\begin{figure}[b]
\centering 
\includegraphics[width=\figmaxwidth]{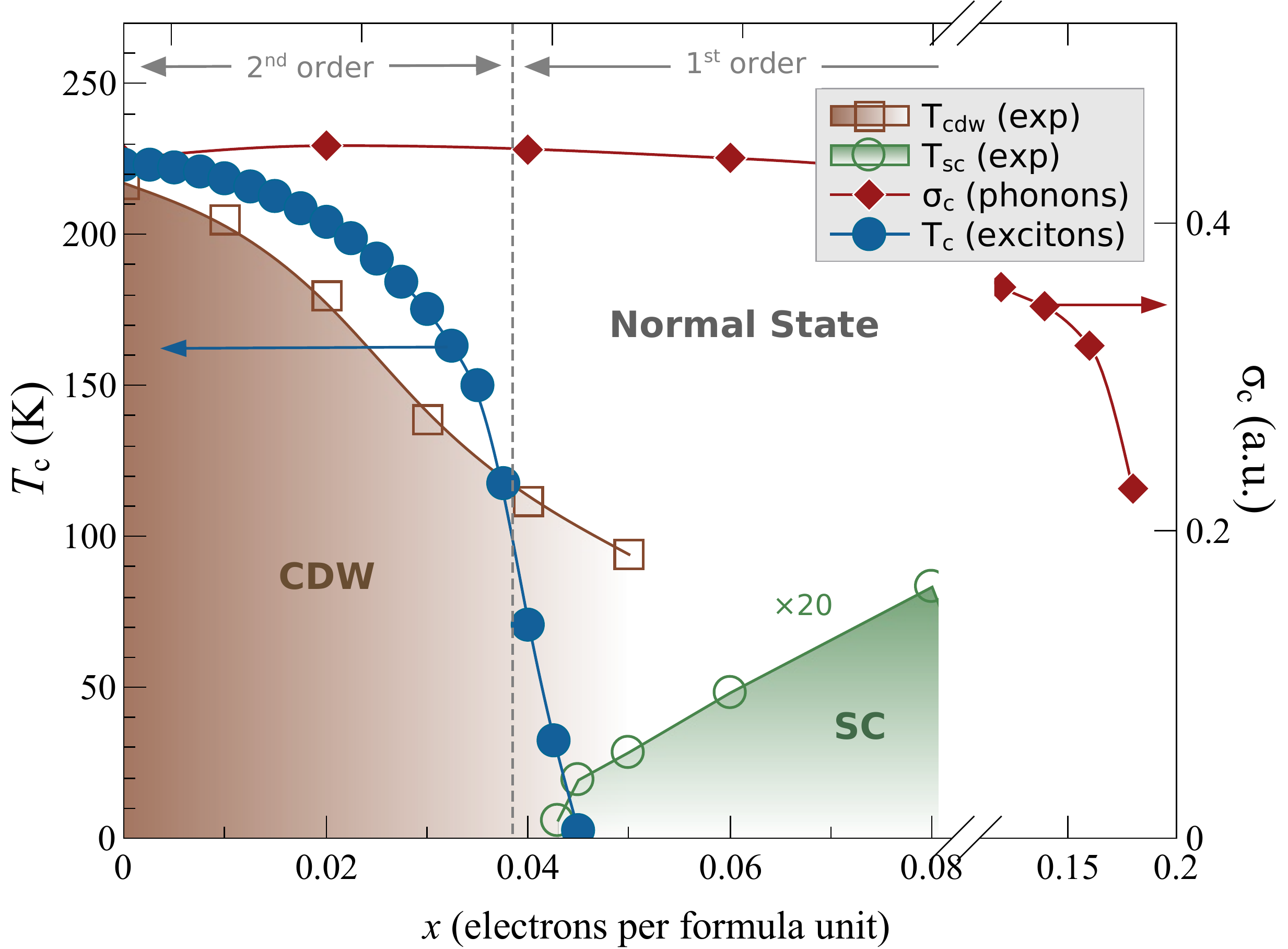}
\caption{
Filled circles represent $\Tc(x)$ obtained from the self-consistent solution 
of the excitonic instability. The strength of the coupling is the only 
adjustable parameter, and was fixed at $V \teq 450$\,meV to yield $\Tc(0) \teq 
220$\,K. 
The diamonds show the critical smearing parameter ($\sigma_c$) above which 
the phonon instability disappears (note the break in the horizontal axis). 
Experimental data for $\Tc$ (open squares) and $\Tsc$ (open circles) 
were extracted from Ref. \onlinecite{Morosan2006}.
}
\label{fig:PhaseDiagram}
\end{figure}

Recently, inelastic X-ray measurements identified a dispersive electronic mode 
compatible with the development of an excitonic condensation at $\Tc$ 
\cite{Kogar2017b}. That excitons and interactions can be important has been 
increasingly better documented by a number of modeling refinements:
Cercellier, Monney, \emph{et al.} showed such mechanism alone could account 
for a number of features observed in the evolution 
of the ARPES spectrum of undoped \TiSe{} \cite{Cercellier2007, Monney2009, 
Monney2011, Monney2015} through the CDW transition;
based on an approximate quasi-1D model, van Wezel \emph{et al.} 
discovered that exciton condensation can enhance the lattice distortion 
\cite{VanWezel2010,VanWezel2010a}.
Hence, the outstanding question is not whether excitonic physics is at play, 
but how much so.

Since the dependence of $\Tc$ on electronic density is well known 
experimentally, we submit that the predicted density dependence of $\Tc$ 
in a description with and without account of the excitonic mechanism should be 
different. As a result, it provides a direct, well defined means to quantify 
the importance of excitonic condensation in the transition to the CDW phase in 
\TiSe. 
Indeed, here we demonstrate that the experimental density dependence of $\Tc$ 
in \CuTiSe{} cannot be captured without explicitly accounting for 
electron-electron interactions and the excitonic instability, as 
summarized in the calculated phase diagram of \Fref{fig:PhaseDiagram}.

\begin{figure}
\centering 
\includegraphics[width=\figmaxwidth]{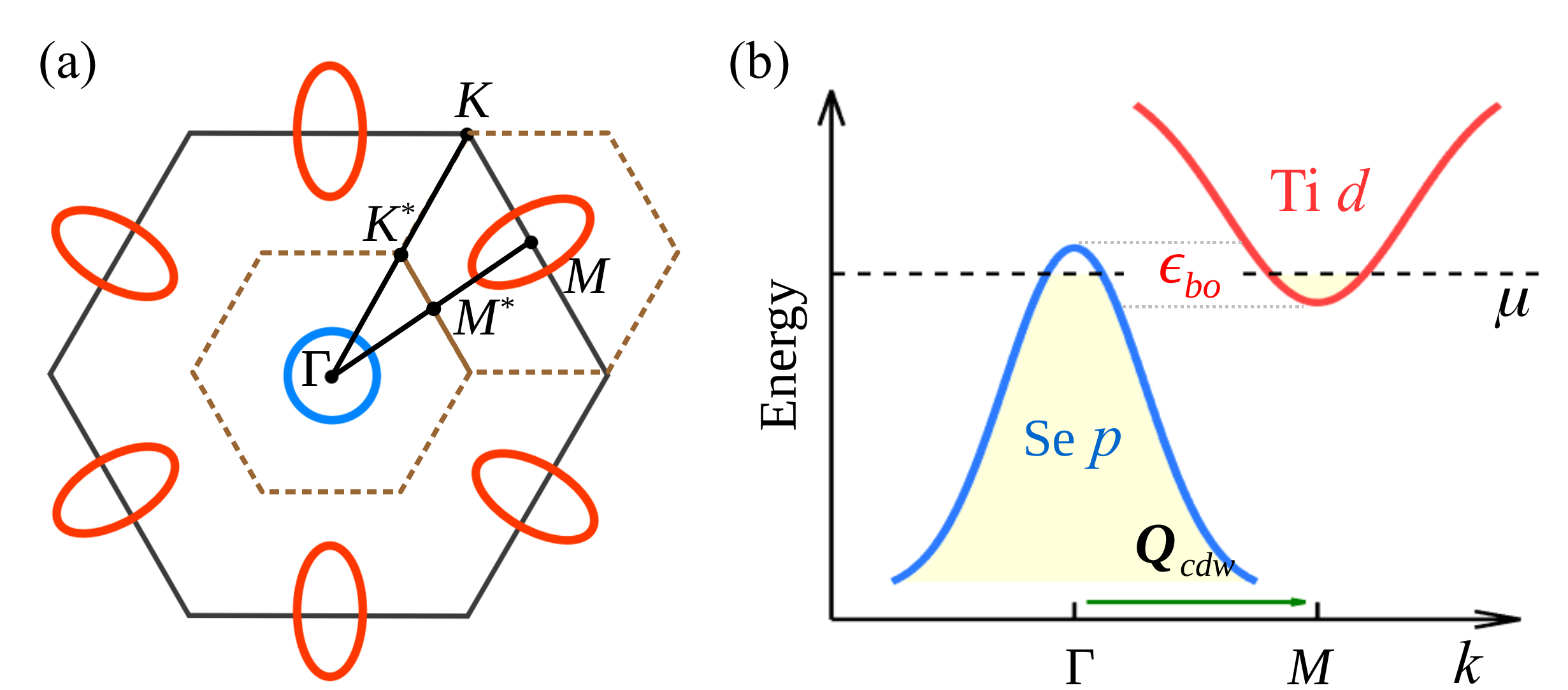}
\caption{
(a) Schematic of the electron ($\Gamma$) and hole ($M$) pockets in the first 
Brillouin zone. The dashed lines highlight the folded zones in the $2\ttimes 
2$ distorted state.
(b) Illustration of the indirect band overlap.
}
\label{fig:band-schematic}
\end{figure}

\begin{figure}
\centering 
\includegraphics[width=\figmaxwidth]{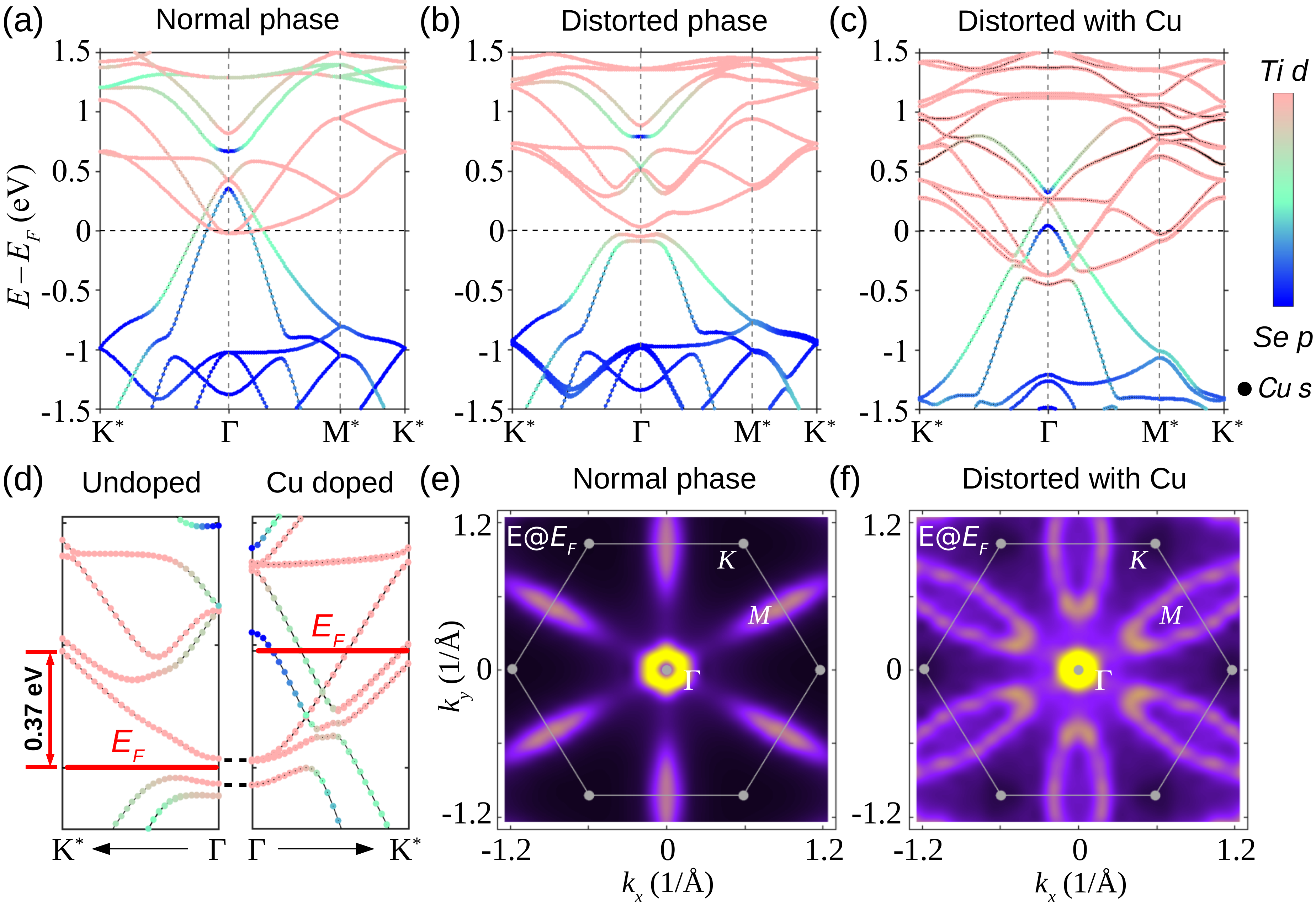} 
\caption{
Band structure obtained with a $2\ttimes2$ cell in the (a) normal clamped, 
(b) relaxed distorted (Ti/Se atoms distort by 0.090/0.029\,{\AA}),
and (c) relaxed configuration with Cu doping. See 
\SFref{7} for the $1\ttimes 1$ unfolded bands. Panel (d) presents a 
side-by-side closeup of (b) and (c). (e) and (f) show the energy contours at 
$E_F$ before and after Cu doping, respectively.
}
\label{fig:CuBands}
\end{figure}

\emph{Excitonic instability}\,---\,%
CDW order is stabilized by intra-layer physics (even in bulk \TiSe, 
\S\  S-IV) which explains the strong similarity of electronic 
and phononic bandstructure changes in monolayer and bulk, as well as their 
doping phase diagram \cite{Morosan2006,Li2015,Chen2016,Goli2012}. Therefore, to 
interrogate whether the excitonic mechanism is able to drive the system through 
a CDW transition in agreement with experiments, we study the \TiSe\ monolayer. 
Although there are two hole pockets [\Fref{fig:CuBands}(a)], we consider only 
the highest one (\S\ S-I.D), similarly to previous studies 
 \cite{Monney2009,Monney2011,Monney2013}. It is modeled as isotropic 
with $\ve_{v\bk} \tequiv \tminus \hbar^2 \bk^2 / 2m_v \tplus
\epsilon_{bo}$, centered at the $\Gamma$ point, while the three electron 
pockets at each $M_i$ point have anisotropic effective masses, 
$\ve_{c\bk,i} \equiv \hbar^2 (\bk\tminus\bM_i)_\perp^2/2m_{c,\perp} \tplus 
\hbar^2  (\bk\tminus\bM_i)_\parallel^2/2m_{c,\parallel} $, as per 
\Fref{fig:band-schematic}. 
When undoped, the chemical potential  ($\mu$) of \TiSe{} is placed near the 
intersection of the conduction and valence pockets, in agreement with the 
folded DFT band structure calculated in an unrelaxed $2\ttimes 2$ superlattice 
[cf. \Fref{fig:CuBands}(a) later], and also tallying with transport experiments 
that reveal both electron and hole carriers in the normal state 
\cite{DiSalvo1976,Morosan2006}.
The band parameters have been extracted by fits to ARPES data in reference
\onlinecite{Chen2015} in the normal state \cite{Parameter}. 
Since the bands strongly renormalize near $E_F$ and CDW fluctuations are likely 
present at $T{\,\gtrsim\,}\Tc$ \cite{Monney2009}, the fitting privileged large 
energy ranges above and below, rather than the close vicinity of $E_F$.
With these, our normal state electron density is $n_e \tsim 4\ttimes 
10^{13}\,\text{cm}^{-2}$, consistent with the experimental Hall data 
\cite{DiSalvo1976} (see also \SFref{4}).

The Hamiltonian comprises these 4 ``bare'' bands and a \emph{direct} Coulomb 
interaction between electrons at the valence and conduction pockets 
\cite{Keldysh1965,Jerome1967,Kohn1967,Monney2009}: 
\begin{align} 
  H & \equiv \sum_{\bk,\sigma} 
    \ve_{v\bk}c_{\bk,\sigma}^{\dagger}c_{\bk,\sigma} 
    + \sum_{\bk,\sigma,i}
    \ve_{c\bk,i}d_{i,\bk,\sigma}^{\dagger}d_{i,\bk,\sigma}
    \nonumber \\
  & 
  +\frac{1}{\mathcal{N}}\sum_{i}\sum_{\bk,\bk',\bq,\sigma,\sigma'} V_{i,\bq}
  c_{\bk+\bq,\sigma}^{\dagger} d_{i,\bk'-\bq,\sigma'}^{\dagger} 
d_{i,\bk',\sigma'} 
  c_{\bk,\sigma}
  ,
  \label{H-def}
\end{align}
Here, $c_{\bk,\sigma}$($d_{i,\bk,\sigma}$) are annihilation operators for 
electrons at the valence ($i$-th conduction) pocket with momentum $\bk$ 
($\bM_i+\bk$) and spin $\sigma$, and $\mathcal{N}$ is the number of unit cells 
of the crystal (for electron pockets at $\bM_i$, $\bk$ represents the momentum 
measured from $\bM_i$). 
The chemical potential $\mu$ is implicit in $\ve_{c\bk/v\bk}$ which 
are measured with respect to it. 
A mean-field decoupling generates the order parameter
\begin{equation} 
  \Delta_{i,\bk,\sigma}(T) \equiv 
  \frac{1}{\mathcal{N}}\sum_{\bk'}V_{i,\bk-\bk'}\langle 
  d_{i,\bk',\sigma}^{\dagger}c_{\bk',\sigma}\rangle
  \label{eq:Delta-def}
\end{equation}
that is directly related to the amplitude of the CDW at $\bk \teq \Qcdw^{(i)}$ 
\cite{SI}.
In view of the $C_3$ symmetry among the three pockets $i$ and the small pocket 
size, we approximate $V_{i,\bq}$ and $\Delta_{i,\bk,\sigma}$ to $i$- and 
$\bk$-independent constants. In particular, $\Delta \equiv 
\Delta_{i,\bk,\sigma}$ is the central quantity for our mapping of the 
temperature-doping phase diagram associated with the excitonic instability. It obeys a 
self-consistent equation \cite{SI}, cf. Eq.~(S8), whose solution for different 
$\mu$ yields the transition temperature $\Tc$ to the CDW phase ($\Delta \tne 0$) 
as a function of doping.

\emph{Self-consistent phase diagram}\,---\,%
\Fref{fig:PhaseDiagram} shows the resulting $\Tc$, calculated entirely 
self-consistently at different doping for the first time, and how it 
compares with the experimental transition temperatures [see also \SFref{2}(a)]. 
It can be clearly seen that: (i) the decreasing trend from $x \teq 0$ follows 
very well the experimental behavior until $x \tapprox 0.038$; (ii) the 
calculation predicts $\Tc\,{\to}\,0$ at precisely the doping where the CDW 
changes from commensurate to incommensurate \cite{Kogar2017a} and SC phase 
emerges ($x \tapprox 0.04$); (iii) the transition is of 
2$^\text{nd}$ order until $x\tapprox 0.038$, becoming 1$^{\text{st}}$ order 
afterwards, which correlates with the doping for the onset of 
discommensurations or ICDW observed in recent experiments \cite{Kogar2017a,SI}.
Having set all the bare band parameters from ARPES data as described earlier, 
\emph{our theory of the charge instability depends only on one parameter}: the 
coupling $V$. We set it at 450\,meV to match the calculated $\Tc$ to the 
experimental one at $x\teq 0$. With $V$ thus fixed, the results 
for $\Tc$ at different $x$ shown in \Fref{fig:PhaseDiagram} follow 
\emph{without further parameters adjustment}. At $x \teq 0$ we have 
$\Delta(0)\tapprox 25$\,meV [\SFref{2}(b)], in reasonable agreement (given 
the approximations) with $\tsim 50$\,meV measured in bulk and monolayer 
\cite{Monney2013,Chen2015} after subtracting background fluctuations from 
the latter, as pointed out by Monney \etal\ \cite{Monney2013,Cercellier2007}.

Experimental confirmation of whether this mechanism is critical or not in 
driving the CDW instability in \TiSe{} and related TMDs can be obtained by 
probing $\Tc$ as a function of both electron and hole doping to establish: (i) 
whether an optimal $\Tc$ exists and (ii) whether it correlates with having 
$\mu$ at the pocket intersection.

Note that the absence of nesting implies that the ``renormalized'' electronic 
bands in the CDW phase are only partially gapped \cite{Monney2009} (\SFref{3}). 
This translates into a predicted increase in the resistivity, $\rho(T)$, as 
soon as CDW fluctuations set in at $T\gtrsim\Tc$, but persistence of the 
metallic nature at low temperatures; notably, holes are suppressed below
$\Tc$. All these features tally with measurements of thermal and electronic 
transport across the transition \cite{DiSalvo1976,Morosan2006,SI}. In addition, 
the preservation of partial electron pockets in the excitonic phase provides a 
Fermi sea for the development of SC beyond a threshold doping, and the 
co-existence of SC and CDW order, as seen experimentally \cite{Spera2017}.

These results reveal that the excitonic mechanism is able to capture correctly 
all the key qualitative aspects of the CDW transition and, in addition, account 
quantitatively very well for the experimental doping dependence of $\Tc$. The 
agreement extends to the position of the CDW critical point that is 
predicted here to lie rather close to the experimental onset of the SC dome.

\emph{Band restructuring ab initio}\,---\,%
To obtain an unbiased perspective over the doping dependence of both the 
reconstructed energy bands and phonon spectrum with doping, we carried out 
extensive DFT calculations \cite{Kohan1964_dft} with the projector augmented 
wave method implemented in the Vienna Ab-initio Simulation Package (VASP) 
\cite{paw1999,vasp1996}. 
Electronic calculations used the generalized gradient approximation (GGA) 
\cite{pbe1996_xc} for the exchange-correlation functional and include 
spin-orbit coupling. The force constants were obtained within density 
functional perturbation theory (DFPT) and the phonon dispersions computed with 
the PHONOPY code \cite{DFPT1987_phonopy,phonopy2008}. Details of these 
calculations and methodology are given in the \SI{} \cite{SI}.
Effect of additional carriers in \TiSe{} were investigated with two 
complementary strategies: directly simulating supercells with adsorbed Cu 
and by adding/removing electrons to the unit cell with a neutralizing 
uniform background charge. 

In the high-temperature undistorted phase, \TiSe{} contains two Se $p$-derived 
hole pockets at the $\Gamma$ point slightly overlapping with three Ti 
$d$-derived electron pockets at the $M$ point (\S\  S-I.D.).
As these are related by $\Qcdw$, in a $2\ttimes2$ superlattice representation 
they fold to the $\Gamma$-point of the reduced Brillouin zone, as explicitly 
shown in \Fref{fig:CuBands}(a). The Fermi 
energy ($\Ef$) is slightly below the intersection of electron and hole pockets, 
as required by charge neutrality given the higher number of electron pockets. 
If one freezes the ions, these bands do not hybridize and revert to their 
respective primitive BZ positions in the unfolded band structure 
[cf. \SFref{7}(b)].  

Relaxing the ions yields a distorted ground-state (the PLD), the overlapping 
pockets hybridize at $\Ef$, and a gap appears ($E_\text{g} \teq 
82$\,meV) resulting in an overall lowering of energy. In addition, there is an 
important restructuring of the bands' shape near $\Ef$ as shown in 
\Fref{fig:CuBands}(b) and \SFref{5}; this causes loss of the parabolic 
dispersion towards an inverted Mexican hat profile. 
In DFT, this feature was first observed in calculations only after adding GW 
quasiparticle corrections to the LDA band structure of bulk \TiSe{} 
\cite{Cazzaniga2012}. Its observation here at the GGA level indicates it 
captures the important qualitative details to accurately describe the low 
density pockets in \TiSe{} (we discuss the electronic structure predicted with 
an alternative HSE hybrid functional in the \SI{} and \SFref{6}).
The commensurate $2\ttimes 2$ PLD ground state, the magnitude of the atomic 
displacements, and nontrivial restructuring of energy bands are in substantial 
agreement with experiments. The unfolded band structure shown in \SFref{7}(d) 
exhibits distinct back-folded bands at the $M$ point that retain the nontrivial 
Mexican hat shape, as has been recorded in ARPES 
\cite{Cercellier2007,Chen2015}.

\emph{Doping by Cu intercalation}\,---\,%
We now add Cu atoms to the monolayer and report in \Fref{fig:CuBands}(c) 
the band structure in the reduced Brillouin zone of a fully optimized 
$2\ttimes2$ supercell with two Cu atoms (one above and one below the \TiSe{} 
slab, to preserve the symmetry). 
This visibly increases $\Ef$ and restores the partial overlap between 
the electron and hole bands: at this doping, the system is a semimetal with 
a rigid upward shift of $\Ef$. This is further evidenced by the Fermi contours 
shown in \Frefs{fig:CuBands}(e,f) that shrink at $\Gamma$ and expand at $M$ to 
cover a large area of the BZ. Despite having been computed without 
and with the Cu atoms, these Fermi contours are adiabatically connected, 
similarly to the evolution of Fermi surfaces in the experiments 
\cite{Zhao2007}. 

There are two crucial effects of doping with Cu.
First, inspection of the bands in \Frefs{fig:CuBands}(b--d) shows that it does 
not remove the nontrivial restructuring of the dispersion near the 
electron-hole intersection of the pristine monolayer; \Fref{fig:CuBands}(d) 
emphasizes this observation by placing the undoped and doped band structures 
near $\Ef$ side by side. 
This agrees with STM measurements showing that the gap in the CDW phase of 
\CuTiSe{} appears below $\Ef$ and moves to higher binding energies 
proportionally to the Cu content \cite{Spera2017}.
Second, an analysis of atomic relaxations further reveals that doping nullifies 
the large atomic displacements observed in the distorted state of the undoped 
system and entirely suppresses the PLD (\SFref{8}).  

Note that the concentration of Cu in these supercell calculations is 
extremely high (Cu:Ti$\teq 50\%$) for direct experimental comparison (the Cu 
solubility limit is 11\% \cite{Morosan2006,Wu2007}).
The crucial factor here is that, \emph{despite} such high doping, our results 
provide clear evidence that the leading effect of Cu adsorption is to 
donate carriers to the conduction bands (one electron per Cu). This rigidly 
shifts $\Ef$ without marked modification of the dispersion and one naturally 
expects a more dilute scenario to introduce even less perturbation beyond 
shifting $\Ef$. Therefore, in order to scrutinize in detail the phonon 
instability at experimentally compatible doping (below 10\,\%), we resort to 
the second doping strategy mentioned above, which would otherwise require 
prohibitively large supercells in the DFT and phonon calculations.

\begin{figure}
\centering 
\includegraphics[width=\figmaxwidth]{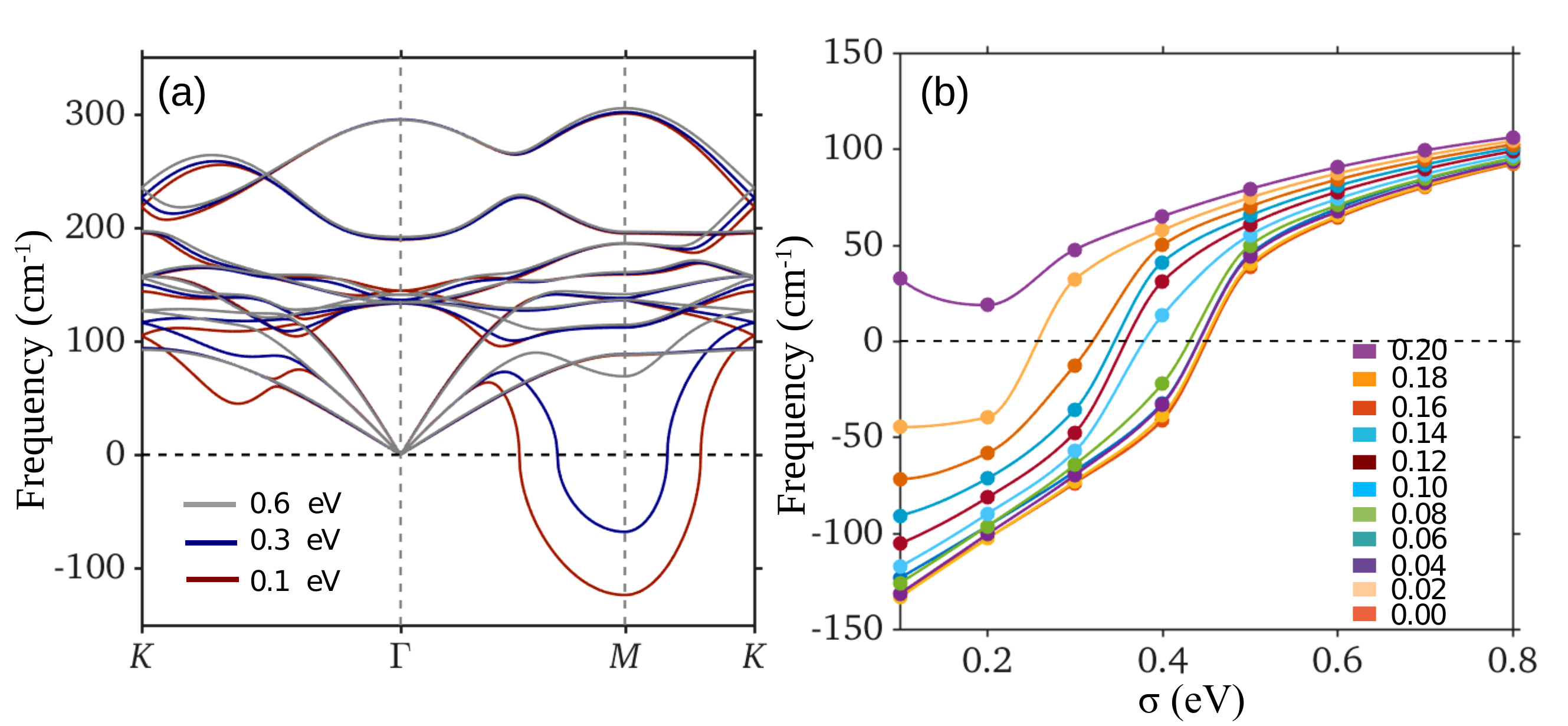}
\caption{
(a) Phonon spectrum of the $1\ttimes 1$ normal phase with different smearing 
parameter $\sigma$ (undoped). 
(b) Evolution of the soft mode frequency at $\bk\teq M$ as a function of 
$\sigma$ (abscissas) and $x$ (legends). Imaginary frequencies are represented 
as negative values.
}
\label{fig:Phonons}
\end{figure}

\emph{Phonon softening ab initio}\,---\,%
\Fref{fig:Phonons}(a) displays the phonon spectrum of \TiSe{} in the normal 
($1\ttimes1$) phase. The qualitative influence of temperature is 
probed by varying the electronic smearing parameter $\sigma$, which is
normally used as a technical tool in the \emph{ab-initio} calculations to 
accelerate the convergence and, in certain circumstances, acquires 
the role of electronic temperature \cite{SI}. A marked dependence of soft modes 
on $\sigma$ is conventionally used to trace \emph{qualitative} changes expected 
to occur in the real phonon spectrum with temperature. At the smallest smearing 
($\sigma \teq 0.1$\,eV), a soft mode with imaginary frequencies (represented as 
negative values) around the $M$ point signals the dynamical instability towards 
the $2\ttimes 2$ PLD observed experimentally below $\Tc$, which is 
complementary to that based on total energy minimization in the $2\ttimes 2$ 
superlattice discussed above.
The fraction of the BZ associated with imaginary frequencies decreases at 
higher $\sigma$ and disappears beyond a critical value $\sigma_c \tapprox
0.45$\,eV (note that only one acoustic mode is sensitive to $\sigma$, as in 
experiments \cite{Holt2001}). This hardening behavior implies that undoped 
\TiSe{} should be stable only above a threshold temperature $\Tc$ because, 
while $\sigma_c$ cannot be directly related to $\Tc$, existence of a finite 
$\sigma_c$ can be safely used to predict a finite $\Tc$ \cite{SI}; this agrees 
with the experimental situation. 

To probe systematically the effect of small uniform doping, we studied the 
phonon spectrum with different concentrations of electrons in the unit cell 
($x$, measured in electrons per formula unit, FU) as outlined above. The range 
of imaginary frequencies gradually decreases as $x$ grows, and the soft mode 
becomes stable above $x_c \tsim 0.18{-}0.20$ [cf. \SFref{9}(b)]. A summary of 
the dependence of $\omega(\bk\teq M)$ on both $x$ and $\sigma$ is shown in 
\Fref{fig:Phonons}(b) for electron doping. A similar progression (not shown) is 
found with hole doping, albeit with a smaller critical density ($\tsim 0.08$ 
holes/FU). Hence, both electron and hole doping suppress the PLD in a \TiSe{} 
monolayer.

The variation of $\sigma_c$ with doping is included in the phase diagram of 
\Fref{fig:PhaseDiagram} for comparison. While our DFPT results correctly 
predict the suppression of the CDW/PLD in doped \TiSe{}, the rate of 
suppression with doping is much smaller than in experiments, resulting in an 
order of magnitude discrepancy between the predicted and experimental $x_c$. 
This conclusion is robust with regards to the smearing method used \cite{SI}.

\emph{Discussion}\,---\,%
We provided the first complete, self-contained theoretical description of 
the influence of both temperature and doping in the CDW phase diagram of 
\TiSe{} in a fully self-consistent way.
The solution of the excitonic instability with doping predicts a phase diagram 
in very good agreement with the experimental $\Tc(x)$. This is significant 
because: our bare band structure is fixed from ARPES data; the single 
interaction parameter $V$ is fixed once in the undoped case; the good agreement 
seen for $\Tc(x)$ follows without any subsequent parameter fitting.
In addition, the electron-phonon coupling can be incorporated straightforwardly 
in this scheme, possibly enhancing the CDW instability \cite{Zenker2013,SI}.

The commensurate nature of the CDW, where both amplitude and phase fluctuating 
modes are gapped \cite{GrunerBook,Lee1993,McMillan1977,Su2017}, and the high 
$\Tc$, generically support relying on a mean-field calculation to describe 
the condensed phase of this problem. However, fluctuations are likely the 
reason for the persistence of the spectral gap in ARPES even above $\Tc$ 
\cite{Cercellier2007,Chen2015}, and for our $\Delta(0) \tminus \Delta(\Tc)$ to 
be $0.5$ [\SFref{2}(b)] of that same difference in experiments for undoped 
\TiSe\ \cite{Cercellier2007, Monney2009, Monney2011, Monney2013}.
The experimental restructuring towards Mexican-hat-shaped bands, with spectral 
transfer affecting only low energies, indicates that the physics is well 
described by our mean-field decoupling scheme.

Although DFT+DFPT implementations capture the electron-phonon coupling and 
some level of electronic correlation, they do not account for the excitonic 
condensation. By not explicitly capturing this physics, the calculation is 
unable to describe the correct degree of phonon softening, especially because 
the very low density places $\Ef$ in the region where the spectrum is 
non-trivially restructured. This sensitivity to electronic interactions tallies 
with previous evidence that DFT-based results for the stability of the PLD and 
renormalized band-structure depend strongly on the exchange and correlation 
functional, the usage of a local or non-local density approximation, and 
quasiparticle corrections \cite{Cazzaniga2012, Olevano2014, Calandra2014a, 
Singh2017, Hellgren2017}.

Our results place the excitonic instability as a decisive element in 
the microscopic description of the CDW/PLD transition, as hinted by recent 
experiments that unveiled hybridized excitonic and phonon modes
\cite{Kogar2017b}. The current ability to map the phase diagram in 
strictly 2D \TiSe{} by gate doping \cite{Li2015} should allow forthcoming 
studies of the yet unexplored hole-doped regime, e.g., whether an optimal $\Tc$ 
correlates with $\Ef$ at the intersection of the electron and hole pockets, as 
predicted here.

\begin{acknowledgments}
We thank Lei Su and A.~H.~Castro~Neto for fruitful discussions. VMP was 
supported by the Singapore Ministry of Education through grant MOE2015-T2-2-059.
Numerical computations were carried out at the HPC facilities of the NUS Centre 
for Advanced 2D Materials, supported by the National Research Foundation of 
Singapore under its Medium-Sized Centre Programme.
\end{acknowledgments}

\bibliographystyle{apsrev4-1}
\bibliography{TiSe2-excins}

\clearpage
\includepdf[pages={{},1,{},2,{},3,{},4,{},5,{},6,{},7,{},8,{},9,{},10,{},11,{},12,{},13,{},14,{},15,{},16,{},17,{},18,{},19,{},20,{},21,{},22,{},23,{},24,{},25,{},26,{},27}]%
{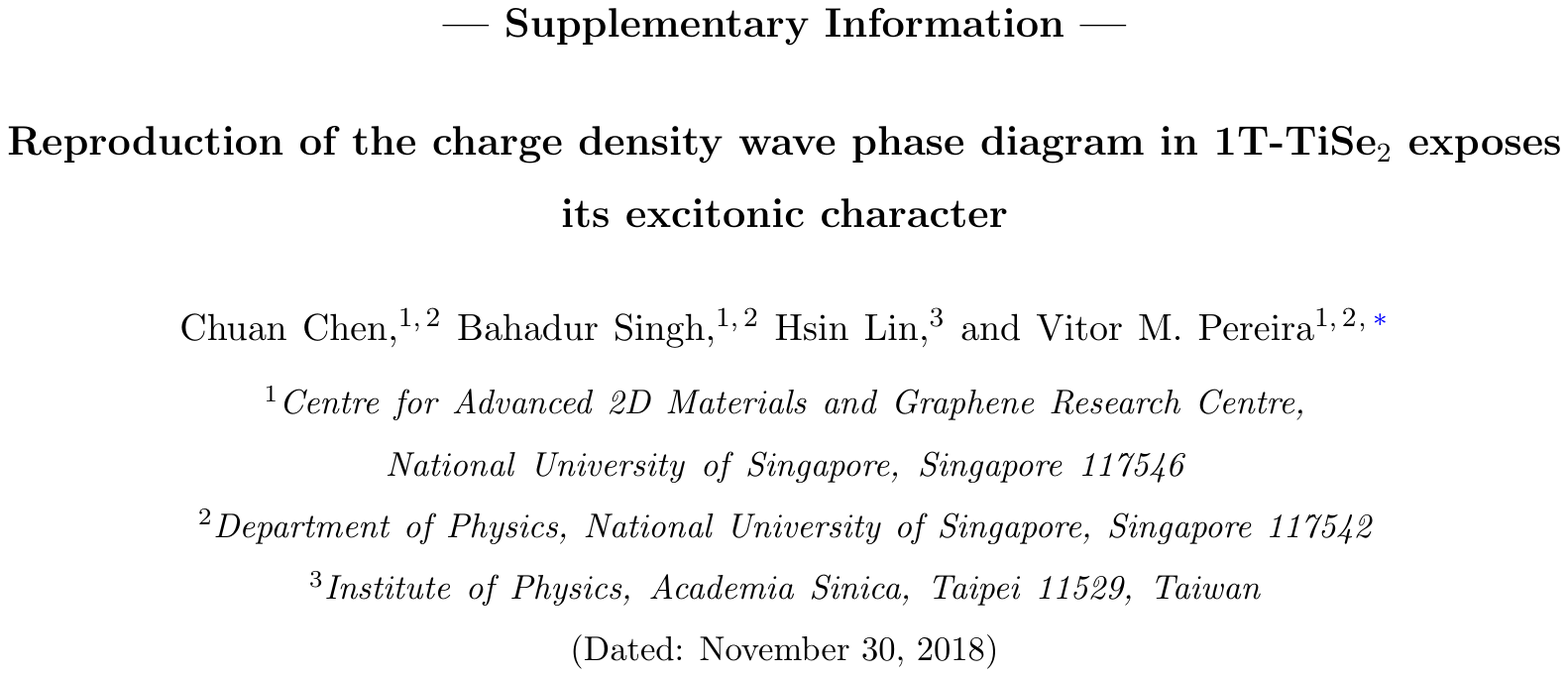}

\end{document}